# OPTICS OPTIONS FOR THE 2015 LHC RUN

M. Giovannozzi, CERN, Geneva, Switzerland

*Abstract*

A review of the possible optics configurations for the 2015 LHC run will be made. The rationale behind the various scenarios will also be presented together with the latest results of the validation studies. Special runs, such as Van der Meer and high-beta, will be discussed too. Finally, the next steps and the related milestones will be discussed with the goal of achieving a consensual decision on the optics configuration to be used for the LHC in the coming weeks.

## POSSIBLE OPTICS CONFIGURATIONS

The overall beam and optical parameters proposed for the 2015 run can be found in Ref. [1], where the rationale behind these choices is discussed in detail. In this paper these values are taken as input and various optical configurations, all compatible with them, are discussed.

The potential changes to the Run I optics can be grouped into three categories depending on their goal, namely:

- Take into account the experience gained during Run I.
- Extend the performance reach of the LHC.
- Prepare for the future.

Of course, a more prudent approach can be applied, considering that the LHC ring underwent important modifications affecting the magnetic circuits. Therefore, sticking to the Run I nominal optics might be a suitable option in view of minimising the risk of additional unforeseen difficulties during the 2015 beam commissioning.

The items presented in this paper as possible optics configurations for the 2015 run have been worked out and presented in detail in Refs. [2-4]. Three options have been devised [3, 4]:

- Option-min: it is the closest configuration to the one used during Run I. Only the change of crossing angle scheme in IR8 [2] is implemented, which is mandatory for operation with 25 ns bunch spacing beams, and the use of all MCBXs for the generation of the crossing and separation schemes. It is worth mentioning that some slight changes have to be made to the squeeze sequences of IR2 [5] (ions [6]) and IR8 [7] to make them compatible with the higher energy with respect to Run I.
- Option-med: with respect to Option-min, the optics of IR4 is modified in order to increase the values of the beta functions at the location of the D3 separation dipole in view of improving the performance of the synchrotron radiation monitor (BSRT). This has also positive side effects on the beam size at several instruments for measuring beam profiles [8, 9] as well as a beneficial impact on the effective strength of the transverse damper [10, 11]. In principle, also the IR6 optics could be upgraded according to what presented in Ref. [12] and assessed in Ref. [13]. This option has been considered not to be necessary.
- Option-max: it consists of an ATS-compatible [14] optics, with a configuration of IR4 fulfilling the requirement of increased beta functions as for Option-med, even if the two solutions are not exactly the same.

It is worth noting that Option-max fulfils all three criteria listed before, as it has been basically tested with pilot beams during Run I [15-19] and it incorporates the required changes in IR4. Moreover, it increases the performance reach by opening the possibility of using flat optics, which provides an interesting boost in performance with longer than nominal bunch length, very large $\beta^*$ values and clean chromatic properties of collision optics, including low spurious dispersion. Finally, it is the HL-LHC baseline optics [20-22] and its implementation in operation would allow gaining experience with such a novel optics concept and it would be therefore beneficial for the upgrade project.

## SOME ADDITIONAL POINTS

There are a number of generic aspects that should be taken into consideration in view of finalising the optics configuration for the 2015 run.

### Tune control

The control of the fractional part of the tune is currently made by means of the phase advance of the local optics of IR1 and 5 [23]. At top energy, the first matched optics of the squeeze sequence performs a variation of phase advance in IR1 and 5 so to change the fractional part of the tune from the injection value of (0.28, 0.31) for the horizontal and vertical plane, respectively, to (0.31, 0.32). This change is performed at constant value of $\beta^*$. During Run I beam losses have been observed during this stage of the squeeze [24], which has been correlated with a too strong orbit change due to the feed down stemming from the quadrupoles that vary the phase advance. A natural solution would be an increase in the duration of such an optics transition. Nevertheless, this would have an adverse impact on the overall duration of the beta-squeeze process, which is certainly not going in the right direction, i.e., of optimising the cycle length for physics.

At the same time, it should not be forgotten that the fractional part of the tune can be controlled via the MQTs [25] with a minimum impact on the beta-beating. Therefore, it is proposed to use these quadrupole correctors to vary the machine tunes. In principle, the optics can be kept constant and the MQTs changed in order to achieve the target tune values for each moment during the cycle. This approach would provide a very flexible means of acting upon the tunes as the duration of the tune transition stage and its location in the LHC magnetic cycle can be changed at will, without any need for additional re-commissioning time.

The most likely choice of the optics to be used could be the one providing as natural tune values the collision ones. The performance in terms of aperture at injection should be carefully checked though [26].

Another aspect of the tune control is the choice of the value of injection tunes. In fact, the nominal working point was meant to cope with relatively large coupling at injection. The experience of Run I showed that coupling is well under control and using the collision tunes at injection does not seem to have any harmful effect as tested in MD studies [27]. Therefore, the flexibility of the proposed solution could be used to start the beam commissioning using the nominal tunes at injection and then to move to the collision tunes at top energy with a transition of the appropriate duration to ensure a gentle effect onto the orbit. Moreover, the tune transition could also overlap with part of the squeeze, but possibly avoiding to perform this gymnastics at too low $\beta^*$ values.

*Special runs*

The 2015 proton run features a non-negligible number of special runs requested by the Experiments. The situation in terms of optics configurations can be summarised as follows [28]:

- LHCf run: the preferred value of $\beta^*$ ranges in the interval between 11 m and 20 m with a negative crossing angle.
- Van der Meer scans: the requests depend on the Experiments. ATLAS, CMS, and Alice aim at a $\beta^*$ value around 20 m, while LHCb requests a $\beta^*$ value in the interval between 30 m and 40 m. The crossing angle should be set to zero.
- High-beta run: the target value of $\beta^*$ is 90 m.

The straightforward approach would consist in combining LHCf and Van der Meer scans in one group, leaving the high-beta run in a second group. This would mean two separate un-squeeze processes.

A first level of improvement could be having a common un-squeeze up to 20 m $\beta^*$. The high-beta un-squeeze would then branch off the common part.

A second level of improvement could be obtained by having a different injection process, in which $\beta^*$ in IR1 and 5 would be around 20 m or 30 m. This would have the advantage of shortening the un-squeeze time required for the high-beta run. Of course, it should be stressed that the reduction of the un-squeeze time would call for the maximum possible value of $\beta^*$ at injection, which should be compatible with aperture constraints. Such constraints, however, might reduce the overall gain in terms of un-squeeze time. On the other hand, this approach would require commissioning a new injection configuration, which would be an overhead for the corresponding physics run. Basically, it has been estimated that such an approach is worth only if the high-beta run is longer than a couple of weeks [29].

To note that another possibility to improve the efficiency would be to perform a combined ramp-and-squeeze [30], but this is not part of the baseline for the beginning of the 2015 run.

*Triplets in IR2 and 8*

Another point to consider is the management of the strength of the triplets in IR2 and 8. It is well known that the constraints from injection and its protection devices impose to run the triplet at higher-than-nominal gradients, i.e., at value of the order of 220 T/m [25] at 7 TeV if the optics is not changed during the ramp. The corresponding circuit rating imposes that the injection optics cannot be kept constant above energies of 6.78 TeV. Hence, beyond this threshold, ramp-and-squeeze gymnastics should be envisaged.

Another constraint is that the triplets' gradient has to be at its nominal value, i.e., 205 T/m, when the beams are put in collision. The reason behind this request is to avoid excessive heat load on the triplets due to the collision debris. This implies that the matching between the injection and the collision strength can be performed either as a separate process from the squeeze proper, the so-called pre-squeeze where the triplets' strength is reduced at constant $\beta^*$ value, or simultaneously with the squeeze process.

The request of operating in collisions with the triplets at their nominal gradient is certainly well justified for the high-luminosity insertions IR1 and 5, but the luminosity for Alice and LHCb is much lower, at the level of 1-$10 \times 10^{29}$ cm$^{-2}$s$^{-1}$ and 4-$6 \times 10^{32}$ cm$^{-2}$s$^{-1}$, respectively, during Run II. Therefore, this point has been raised and a formal statement is expected from the MP3 [31]. A confirmation that a reduction of the triplets' strength is indeed possible would highly simplify the optics changes at least below 6.78 TeV.

## STATUS OF VALIDATION STUDIES

As a follow up of the proposal presented in Ref. [4], the validation of Option-max has been launched, based on the comparison with Option-min of: dynamic aperture (DA) [32], cleaning efficiency, and machine protection [33]. At the same time, the proposed crossing scheme in IR8 has been evaluated in terms of aperture for injection failure scenarios [34].

The detailed numerical simulations of DA including several configurations, i.e., with or without beam-beam effects, with or without Landau octupoles, did not show any relevant difference between Option-min and Option-max. Also, the situation of beam aperture at injection for the new crossing scheme is compatible with the requirements.

On the other hand, the simulations of the cleaning efficiency did reveal differences between the two optics configurations. Moreover, the situation in terms of machine protection is made worse for Option-max by the imposed phase advance between the dump kicker and the TCT for Beam 2. To mitigate this, a certain reduction in β* reach should be accepted. All in all, the LMC decided that further clarification of the actual cleaning performance of Option-max should be carried out with dedicated measurements in 2015 and that this option would not have been the one for the initial beam commissioning. Given the relative comparison, the validation process essentially gave the green light to Option-min as suitable optics configuration for 2015, with the need of some further verifications for the case with β* =80 cm. Nonetheless, the LMC asked to proceed with the validation of Option-med in view of the benefits for instrumentation and transverse damper.

## NEXT STEPS

The forthcoming weeks, four to eight, will see the optics activities focusing on two main fronts.

### *Validation of Optics-med*

The validation task will be performed by assessing the performance in terms of DA, cleaning efficiency, and machine protection. For Beam 1, only the IR4 optics has changed and at constant IR phase advance. On the other hand, for Beam 2 the change of IR4 optics is also accompanied by a change of IR phase advance, which has been compensated in IR8 [35]. While the overall machine phase advance is kept constant, the phase relation between locations far away in the ring is changed with respect to Option-min. In particular, between IP1 and 5 the phase advance is different with respect to the nominal optics, thus requiring a careful check in particular in terms of beam-beam effects.

### *Preparation of optics database*

The validation activities require preparation of the LHC optics database, which is also needed for the generation of the settings required for LHC operation in 2015.

The repository is maintained under afs, and a number of changes are in any case needed, such as the preparation of a new sequence extracted from the layout database, which is compatible with the actual configuration of the LHC ring after LS1, in particular including the non-conformities found [36]. Moreover, the overall structure of the directories will be reviewed taking into account the experience gained during Run I, in particular the need to simply the structure of the various directories and the naming convention used for the strength files, in view of making easier assembling the machine configuration when starting from the configuration of the individual insertions.

In addition, one should not forget that Option-med is built upon Option-min configuration, by adding the specific configuration for IR4 and IR8 (for Beam 2). Therefore, the configuration files for Option-min have to be generated, starting from the clean-up of the nominal optics files.

In particular, the squeeze of IR1, 2, and 5 has to be adapted to avoid that some trim quadrupoles running out of strength. The crossing schemes have to be reviewed by spreading the strength on the three MCBXs. The new crossing scheme in IR8 has to be implemented.

## CONCLUSIONS

After the astonishing performance of the LHC during Run I, the machine underwent an important consolidation during LS1. Several optics options are at hand for Run II and in this paper the three main configurations for 2015 have been presented and discussed in detail.

These configurations differ for the amount of changes with respect to the nominal LHC optics as described in the LHC design report.

A number of more general aspects has been discussed, whose implementation does not depend on the final choice of the optics.

Validations studies are in progress to assess the suitability of each of the available configurations. The first step has been a direct comparison of Option-min and Option-max, which resulted in the decision of not starting the beam commissioning in 2015 with Option-max and to perform additional checks with beam during dedicated beam study periods. It is clear that in the meantime additional efforts will be devoted to the further analysis and understanding of the behaviour of option-max.

The next step will consist of assessing the performance of Option-med, which will then be presented at the LMC for approval as optics configuration for the 2015 run. In case of doubts Option-min will remain as fall back solution for the beginning of Run II.

## ACKNOWLEDGEMENTS

Several discussions and invaluable input from G. Arduini, R. Bruce, H. Burkhardt, B. Dehning, B. Gorini, W. Höfle, J. Jowett, M. Lamont, R. de Maria, E. Meschi, D. Mirarchi, S. Redaelli, T. Risselada, R. Tomas, and J. Wenninger are warmly and gratefully acknowledged. S. Fartoukh deserves a special thank as he has been the driving force of most of the activities covered in this paper.